\documentclass[aps,pre,twocolumn,galleyproof]{revtex4}

\usepackage{rotating}
\usepackage{graphicx}
\usepackage{amssymb,amsfonts,amsmath}
\usepackage{dcolumn}
\usepackage{natbib}
\usepackage[all]{xy}
\usepackage{booktabs}
\usepackage{multirow}
\usepackage{color}
\usepackage{amsmath,epsfig}
\usepackage{url}

\begin{document}

\title{Sicily and the development of Econophysics: the pioneering work of Ettore Majorana and the Econophysics Workshop in Palermo}

\author{Rosario N. Mantegna}
\affiliation{Center for Network Science and Department of Economics, Central European University, N\'ador 9, H-1051, Budapest, Hungary}
\affiliation{Dipartimento di Fisica e Chimica, Universit\`a di Palermo, Viale delle Scienze, Ed. 18, I-90128, Palermo, Italy}


\begin{abstract}
Sicily has played an important role in the development of the new research area named "Econophysics". In fact some key ideas supporting this new hybrid discipline were originally formulated in a pioneering work of the Sicilian born physicist Ettore Majorana. The article he wrote was entitled "The value of statistical laws in physics and social sciences". I will discuss its origin and history that has been recently discovered in the study of Stefano Roncoroni. This recent study documents the true reasons and motivations that triggered the pioneering work of Majorana. It also shows that the description of this work provided by Edoardo Amaldi was shallow and misleading. In the second part of the talk I will recollect the first years of development of econophysics and in particular the role of the "International Workshop on Econophysics and Statistical Finance" held in Palermo on 28-30 September 1998 and the setting in 1999 of the "Observatory of Complex Systems" the research group on Econophysics of Palermo University and Istituto Nazionale di Fisica della Materia.
\end{abstract}
\maketitle


\section{Introduction}
Econophysics \cite{MantegnaStanley} is a hybrid discipline at the interface between physics and economics that has developed since 1990s. Econophysics uses methods and tools of statistical physics, complexity science, network science and data mining in the analysis and modeling of economic and social systems. Similarly to the development of several new disciplines the development of Econophysics was gradual and was characterized by the selection of scientific problems pursued by a group of pioneers, by the editorial policy of academic journals, by the foundation of new journals, and by the organization of Workshops and Congresses that have connected the first research community working in the new discipline. During the setting of a new research community, the identity of the community is originally shaped by the cultural background of the first participants and by some cultural heritage that assumes the role of a pioneering view.
The paper of Majorana on ``The value of statistical laws in physics and in social science'' \cite{Majorana1942,Majorana2005,Bassani2006} was the most influential historical paper that was reconsidered and discussed during the first years of development of Econophysics. We will show that the origin of this paper are rooted in Sicily. More generally Sicily, the biggest island of Italy, was contributing to the development of this new discipline under several respects. In fact, the first Econophysics paper \cite{Mantegna1991} presents as one of the affiliations of the author the University of Palermo. Sicily was also the location for one of the first meetings on Econophysics. This was the ``International Workshop on Econophysics and Statistical Finance'' held at University of Palermo, Italy in 1998. This Workshop is the first Workshop on Econophysics that has published proceedings \cite{Mantegna1999}. Finally, one of the first Econophysics research groups was set in Palermo, Italy in 1999 when a group of young researchers established the ``Observatory of Complex Systems'' that specialized during the first years in the modeling of high-frequency financial data and in the early investigations of financial correlations.
In this article I will report on some recent discoveries \cite{Roncoroni2012} about the true reasons that motivated Ettore Majorana to wrote his paper on the value of statistical laws. I will also recall the 1997-1998 period of the first Econophysics meetings. Specifically, I will focus on the Budapest meeting of July 1997, on the Rome (March 1998), and on the Palermo (September 1998) meetings. A brief recollection about the first years of activity of the ``Observatory of Complex Systems'' will also be provided.

\section{Majorana article on ``The value of statistical laws in physics and in social science''}\label{Sec2}
The history about the article of Ettore Majorana on ``The value of statistical laws in physics and in social science'' is quite intriguing. During the years the information available about his origin, nature, and scope has changed thanks to new studies of the original documents. Ettore Majorana was not a prolific author. He published 9 articles before his disappearance and a tenth article was published in 1942, thanks to the interest of his former friend Giovanni Gentile Jr.. 

As for the majority of Majorana's papers, the paper was originally published in Italian. Also for this reason the paper was not accessible to a large readership until the paper was translated in English in 2005 \cite{Majorana2005,Bassani2006}. However, the original language was not the only reason why the paper was not know to a wide readership. 
The article presents a physicist's point of view about the value of statistical laws in physics and social sciences to scholars of different disciplines such as sociology or economics. Ettore Majorana is focusing on the observation that quantum mechanics has forced physicists to recognize that laws as basic as the radioactive decay of a single atom must contain a statistical character. Majorana then conclude that the presence of an inevitable statistical character in basic physics laws shows that there is an ``essential analogy between physics and social sciences, between which it turned out an identity of value and method''.
In fact, before the setting of Econophysics, this paper was considered by most of the leading Italian physicists as a rather peculiar one. For example, in  his biography of Ettore Majorana, Edoardo Amaldi \cite{Amaldi1966} just wrote a single sentence  \cite{note1} without any scientific comment on it in a biographical and scientific note of 49 pages. Even today, Majorana's point of view might perhaps still be rather unpopular among traditional physicists, in spite of more than 70 years of quantum mechanics and after some major breakthrough in the fields of critical phenomena and chaos theory.
Until a few years ago there were not many details on the origins, motivations and writing procedure of the article on the value of statistical laws was. The only public source about the writing process was Giovanni Gentile Jr that wrote, in a note of the published version of the paper in Scientia \cite{Majorana1942}, the following text: {\it This article of ETTORE MAJORANA - the great theoretical physicist of Naples University who was missing on March 25, 1938 - was originally written for a sociology journal. It was not published perhaps due to the reticence that the author had in interacting with others. Reticence that convinced him to put inside a drawer important papers too often. This article has been conserved by the dedicated care of his brother and it is presented here not only for the intrinsic interest of the topic but above all because it shows us one aspect of the rich personality of MAJORANA that so much impressed people that knew him. Thinker with a sharp realistic sense and with an extremely critics but not skeptic mind. He takes here a clear position concerning the debated problem of the statistical value of basic physics laws. This aspect considered by several scholars a defect similar to a charge of indeterminism in the evolution of nature; it is indeed for MAJORANA a reason to claim the intrinsic importance of the statistical method. Up to now this method has been applied only to social sciences and in the new interpretation of physics laws it fully recovers its original meaning. GIOVANNI GENTILE jr., 1942} \cite{note2}. 

The recent study of Stefano Roncoroni \cite{Roncoroni2012} shows that the origin, motivations, and goal of Majorana's paper were completely different from what was stated by Giovanni Gentile Jr. in the Scientia short presentation of the paper. The documents found by Roncoroni at the ``Fondo Majorana'' of the Regional Library Gianbattista Caruso of Catania, Italy, show that the motivation for the writing of the manuscript originated within the Majorana family. In fact, the article was requested to Ettore from his uncle Giuseppe Majorana. The article was originally intended to be included in a volume celebrating the academic achievements of his uncle just at the period of his retirement. Giuseppe Majorana was a professor of Economics at Catania University with research interests in theoretical and applied statistics \cite{Majorana1889a,Majorana1889b}. The request of the paper was done by Giuseppe Majorana with a letter that was received by Ettore in January 1936. The letter of acceptance from Ettore Majorana to his uncle is dated January 30th, 1936. Most probably Ettore started to work on the article quite soon after his response and completed it by March 4th, 1936. In fact the first ver-sion of the manuscript (essentially coincident with the version we know from the Scientia publication) was sent to Giuseppe by mail on March 4th, 1936. Differently from what stated in Amaldi's biography and in the short text of Gentile, these documents show that in 1936 Ettore was interested into physics, and reflecting on the impact of the new concepts of quantum mechanics on other disciplines. Moreover, he had no reluctance to publish what he wrote and was also quite fast in finalizing his contribution \cite{note3}.

\section{The first Econophysics meetings}\label{Sec3}
The first Econophysics meeting was organized by Imre Csabai, J\'anos Kert\'esz and Imre Kondor in Budapest on July 21-27, 1997. The meeting was called ``Workshop on Econophysics''. The opening talk entitled ``Can statistical physics contribute to the science of economics?'' was delivered by H. Eugene Stanley. The program listed 26 one-hour talks \cite{note4}. The workshop was attended by 60 participants \cite{note5}. The second Econophysics meeting was organized by Maurizio Serva, Angelo Vulpiani and Yi-Cheng Zhang in Rome on March 12-13, 1998. This meeting was called ``Informal Meeting on Statistical Finance''. The opening talk entitled ``Power-laws, information and hierarchy in financial markets'' was delivered by Rosario Mantegna. In this meeting there were 16 talks \cite{note6}. The list of participants to the Rome meeting has 32 names. These two meetings were followed by the ``International Workshop on Econophysics and Statistical Finance'' that was organized by Rosario Mantegna in Palermo on September 28-30, 1998. The Palermo meeting was the first Econophysics meeting with a Scientific Committee. The Scientific Committee was composed by Erik Aurell, Jean-Philippe Bouchaud, Werner Ebeling, Imre Kondor, H. Eugene Stanley, Hideki Takayasu, Angelo Vulpiani, and Yi-Cheng Zhang. It should be noted that the name of the Workshop is citing both the term Econophysics and the term Statistical Finance that were previously used for the Budapest meeting and for the Rome meeting respectively. The reason for this choice was that the Scientific Committee was at that time still not unanimous about the use of the recently coined term Econophysics and some of the members expressed a strong preference for the term Statistical Finance. The chosen name was a compromise between the two lines of thought. The opening lecture entitled ``Can statistical physics contribute to the science of economics?'' was again delivered by H. Eugene Stanley. The meeting had a program with 32 talks \cite{note7} and the number of participants were 53. The Palermo meeting was also the first Econophysics meeting \cite{note8} with published proceedings \cite{Mantegna1999}.

\section{The Observatory of Complex Systems}\label{Sec4}
In 1999 Giovanni Bonanno, Fabrizio Lillo and Rosario N. Mantegna established a research group at Palermo University. Giovanni Bonanno and Fabrizio Lillo had a fellowship from the Istituto Nazionale per la Fisica della Materia (INFM) and Rosario N. Mantegna was an associate professor of applied physics at Palermo University. They called their research group the ``Observatory of Complex Systems''. The research group was part of the Dipartimento di Energetica ed Applicazioni di Fisica, a University Department of Engineers and Physicists. The fundings for the research projects were obtained through INFM and national and international research funding agencies. The INFM research projects were selected through a peer review procedure. A few months later in 2000, Salvatore Miccich\`e joined the research group thanks to another INFM fellowship. The choice of the name of the research group was motivated by the consideration that the availability of large amount of data would have soon changed the research style of many disciplines including finance and economics. In other words, it was foresee that many research approaches would have moved from the customary approach of many disciplines limiting as much as possible the object of the empirical investigation to an observational approach where many processes are observed simultaneously. The Observatory of Complex Systems collected a large database of high frequency financial data and performed research on correlation based networks and financial microstructure.

\section{Conclusions}\label{Sec5}
Sicily has therefore played a multiple role in the development of Econophysics. The origin of the paper of Ettore Majorana on ``The value of statistical laws in physics and in social science'' is rooted in Sicily. In fact, the writing of this article was requested to Ettore from his uncle Giuseppe to be included in a volume celebrating his academic achievements at the time of his retirement, and Giuseppe Majorana was a professor of Economics at Catania University. The third meeting on Econophysics tooks place at Palermo University in 1998. This was also the first meeting on Econophysics with published Proceedings. Moreover, one of the first research groups fully devoted to econophysics research was established in Palermo in 1999.


\begin{thebibliography}{}
\bibitem{note8}
The organizers of the Budapest meeting had also a plan to realize a volume of proceedings. Unfortunately, these Proceedings were not published.
\end{thebibliography}
\end{document}